\documentclass[twocolumn,english,aps,floatfix,superscriptaddress,showpacs,amsfonts,amssymb,superscriptaddress]{revtex4}
\usepackage{color}

\usepackage{graphicx}
\usepackage{amsmath}
\usepackage{latexsym}
\usepackage{epstopdf}
\usepackage{amsmath}
\usepackage{amssymb,amsmath}
\usepackage{multirow}
\usepackage{mathtools}
\newcommand{\beq}{\begin{equation}}
\newcommand{\eeq}{\end{equation}}

\newcommand{\be}{\begin{equation}}
\newcommand{\ee}{\end{equation}}

\begin{document}

\title{Quantum quench in long-range field theories}

\author{M.~A.~Rajabpour}
\affiliation{ Instituto de F\'isica, Universidade Federal Fluminense, Av. Gal. Milton Tavares de Souza s/n, Gragoat\'a, 24210-346, Niter\'oi, RJ, Brazil}
\author{S. Sotiriadis}

\affiliation{ Dipartimento di Fisica dell' Universit\'a di Pisa and INFN, Sezione Pisa 56127 Pisa, Italy}
\date{\today{}}

\begin{abstract}
We study equilibration properties of observables in long-range field theories after the mass quench in $d=1,2$ and $3$ dimensions.
We classify the regimes where we expect equilibration or its absence in different dimensions. In addition we study the effect of the initial 
state in the equilibration properties of our system. 
In the case of free field theories  we show that whenever we have equilibration the long-time limit of correlations can be 
described by the Generalized Gibbs Ensemble. We prove that all integrals of motions in our system are non-local.
\end{abstract}
\maketitle
\section{Introduction}

In recent years due to interesting experiments, \emph{long-range} interacting systems moved into the focus of research. 
Using ultracold atoms and trapped ions techniques many different quantum long-range spin models were engineered \cite{Experiments1} and their time dependent properties were investigated \cite{Experiments2}. 
There has been also considerable progress in the theoretical understanding of quantum long-range systems \cite{theory1}. Different properties such as entanglement entropy
 \cite{theoryentanglement1,theoryentanglement2}
dynamics of entanglement \cite{theory dynamic of entanglement} and equilibration properties \cite{equilibration,Eisert2013,Kastner2014,Levitov,Yuzbashian,Gurarie} 
were studied extensively. 
One of the interesting features of long-range systems is the absence of \emph{Lieb-Robinson velocity} \cite{Generalized Lieb-robinson}. Based on the Lieb-Robinson
 theorem for short-range systems, the effect of a perturbation in part of the system at a given time becomes exponentially small with the distance 
outside a region called the \emph{causal region} \cite{LR72}. This effect which is usually called quasi-locality helps to define a horizon-like 
region inside which the effect of the perturbation is non-zero while outside it is exponentially small. This picture is consistent with the presence 
of a finite Lieb-Robinson velocity that plays the same role as the speed of light in Lorentz invariant field theories. In the presence of long-range 
interactions the above picture is no-longer true. Although in particular regimes one can still find a generalized Lieb-Robinson bound \cite{Generalized Lieb-robinson} for these systems the general form of the causal region is no longer cone-like, as when there exists a maximum velocity. 
The causal region in these cases grows logarithmically for large distances \cite{Generalized Lieb-robinson}. 

One of the important features  of long-range systems is the presence of a power-law like dispersion relation, e.g. $\omega(k)=|k|^{\alpha/2}$ with $0<\alpha<2$. For 
example starting from a system of hard-core bosons with hopping ${|i-j|^{-(d+\alpha)}}$ one arrives at the dispersion relation 
$\omega(k)=|k|^{\alpha/2}$ in momentum space \cite{Cazalila20142}. As we will see soon, the same is true also for coupled harmonic
 oscillators and it is widely believed that this should be the case also for long-range spin systems \cite{DB01}. Although, as we mentioned, 
the long-range systems have been already studied with many experimental and numerical methods, a direct study of a bosonic system with the simple 
dispersion relation $\omega^2(k)=|k|^{\alpha}+m^{\alpha}$ has initiated just recently in the context of entanglement entropy \cite{theoryentanglement2}. The 
interest on this model is two fold: firstly this is among the few systems where one can do analytical calculations and secondly despite
its simplicity it captures most of the interesting features of the more complicated long-range interacting systems. 

Having these two motivations in mind we are interested in studying the evolution of the two-point correlation functions after a sudden change of 
some parameter in a long-range system of harmonic oscillators. Such out-of-equilibrium problems belong to the general category of 
\emph{quantum quenches} \cite{CC06,CardyCalabrese2007,c-06,caz,Rigol,CE-08,bs-08,r-09,r-09a,CS2010,IFTQQ,CEF,eef-12,se-12,rs-12,CE-10,BSE14}. 
We first define our quench protocol and then we demonstrate  different properties emerging from the dynamics. In particular we focus on two 
important features: the presence of horizon effect and the occurrence of equilibration. We show in which circumstances we expect a causal region 
or horizon and we study in detail the regimes where we expect equilibration of correlations. We also discuss the physical interpretation of this
equilibration in the so-called \emph{semiclassical approximation}\cite{CC06,CardyCalabrese2007,CC05,RI11,FC08,Ev13}. Next we verify that, in the 
regimes where equilibration occurs, the long-time properties can be described by a statistical ensemble known as Generalized Gibbs Ensemble (GGE) 
\cite{Rigol,SPR11,R14,ck-14,CSC13a,KCC-13,RS14,SC14,BKC,IFTQQ,andrei,PFE,GGE LL,ce-13,stm-13,fcec-13,GGE failure}. The latter is deduced by an entropy 
maximization principle, subject to all constraints imposed by the conservation laws of the free dynamics (whose number is only extensive, in contrast
to the dimension of the Hilbert space, which increases exponentially with the system size). We also point out that there is no way to express 
the GGE in terms of \emph{local} integrals of motion, in contrast to the usual case \cite{CEF,Fagotti2013,f-14}. Finally we also discuss how turning on the
interactions can modify our results.

\section{Definition of the model}

The system that we are interested in is a system of coupled harmonic oscillators that are described in momentum space by the following quadratic Hamiltonian
\begin{equation} \label{Quadratic Hamiltonian}
H=\sum_k \left(\frac{1}{2}\pi_k\pi_{-k}+\frac{1}{2}\omega^2(k)\phi_k\phi_{-k}\right)
\end{equation}
In order to study the effect of long-range couplings in the evolution and distinguish from short-range systems, 
we will consider two different types of dispersion relations, one of which corresponds to long-range systems (type-1) 
and the other to short-range (type-2), while both exhibiting the same large momentum scaling (Fig.~\ref{fig.SC})
\begin{eqnarray} \label{Dispersion relation1}
&\text{Type-1: } \qquad \omega_1(k)=&\sqrt{|k|^{\alpha}+m^\alpha},\\
\label{Dispersion relation2}
&\text{Type-2: } \qquad \omega_2(k)=&(k^2+m^2)^{\frac{\alpha}{4}}.
\end{eqnarray}
The parameter $m$ will be called the ``mass'' parameter and $\alpha$ is a real parameter that controls 
the long-range scaling of the couplings in type-1. When $m=0$ both types are identical and give the massless 
dispersion relation $\omega(k)=|k|^\alpha$. Also they both reduce to the simple local relativistic dispersion relation for $\alpha=2$. 

For $0<\alpha<2$, type-1 dispersion relation corresponds to long-range couplings in real space. Indeed the Hamiltonian in real space has, in this case, the form
\be \label{Quadratic Hamiltonian real space omega1}
H=\frac{1}{2}\sum_r\pi^2(r)+\frac{1}{2}\sum_{r,r'}\phi(r)\Big{(}\frac{1}{|r-r'|^{d+\alpha}}+m^{\alpha}\delta_{rr'}\Big{)}\phi(r')
\ee

On the other hand, type-2 dispersion relation corresponds to the following Hamiltonian
\begin{eqnarray} \label{Quadratic Hamiltonian real space omega1}
H=\frac{1}{2}\sum_r\pi^2(r)+\hspace{5cm}\nonumber\\
\frac{1}{2}\sum_{r,r'} \left(\frac{m}{|r-r'|}\right)^{\frac{d+\alpha}{2}} K_{\frac{d+\alpha}{2}}(m|r-r'|) \, \phi(r)\phi(r') ,
\end{eqnarray}
where $K$ is the modified Bessel function. Even though this Hamiltonian is non-local in the strict sense (it involves a double space integral), 
it is actually short-range because the couplings decay exponentially with the distance.

Another important point is that in type-1 case for $\alpha>2$ the 
couplings decay like $\sum_{r,r'}\phi(r)\frac{\bigtriangledown^2_{r'}}{|r-r'|^{d+\alpha-2}}\phi(r')$. 
In principle when one studies, for example, spin systems with long-range couplings, even if we do 
not include short-range couplings explicitly at the beginning, they will appear after renormalization anyway, so if we accept the presence of an additional local interaction $\alpha=2$ in all calculations, then one can argue that the interactions with $\alpha>2$ decay fast enough so that one can ignore them in favour of the local interactions. In other words the interactions for $\alpha>2$ are effectively local. We will comment a bit more about this issue in section \ref{sec:VI}.

In the massless case $m=0$, the ground state two-point correlation function is given by
\begin{align} \label{massless propagator}
G(r)=\int \frac{d^d k}{(2\pi)^d} e^{i\mathbf{k}.\mathbf{r}} \frac{1}{2\omega(k)} \sim r^{-d+\frac{\alpha}{2}}.
\end{align}
i.e. at large distances it decays as a power-law.

The massive case is more complex, as the ground state correlations exhibit qualitatively different large-distance behavior in the two different types of dispersion relations. For type-1, the asymptotic behavior of the integral for large $r$ gives
\be \label{propagator omega1 assymptotic}
G(r)\sim \frac{1}{r^{\alpha+d}}\;,\qquad m\neq0, \, 0<\alpha<2.
\ee

The situation is different for type-2 dispersion relation: the integral can be done explicitly and the result is
\be \label{propagator omega2}
G(r)\propto r^{\frac\alpha4-\frac d 2}K_{{(\alpha/2-d)}/{2}}(r),
\ee
where $K$, as before, is the modified Bessel function. For large distances the above expression behaves like
\be \label{propagator omega2 assymptotic}
G(r)\sim r^{\frac\alpha 4 - \frac{d+1}{2}} e^{-mr} \;,\qquad m\neq0, \, 0<\alpha\leq 2.
\ee

It is interesting that the ``massive'' type-1 propagator does not decay exponentially, but as a power law instead. 
The difference is clear when we consider the analytical properties of the integrands in the complex $k$-plane. For type-1 
the integrand has branch-cuts along the imaginary axis starting from $k=0$ (since $|k|=\sqrt{k^2}$), while for type-2 it has 
branch cuts starting from $k=\pm im$ and is analytic at $k=0$ (Fig.~\ref{fig.SC}). In order to extract the asymptotic behavior of 
the integrals for large $r$ we can deform the integration contour so as to wrap the branch cuts. Then in type-2 case we get an exponential 
factor $e^{-mr}$ that is absent in type-1 case. This is the main difference between $\omega_1(k)$ and $\omega_2(k)$ and provides the option 
to investigate the effect of short/long-range couplings in the post-quench evolution. The above mathematical treatment 
will be used also later in the calculation of the post-quench correlations.

\subsection{Generalized Lieb-Robinson bound}
In this subsection we would like to classify  different regimes where a Lieb-Robinson bound or its generalized form exists. 
We first present the Lieb-Robinson theorem \cite{LR72}: consider the Hamiltonian $H_{\Lambda} =\sum_{X \subset\Lambda} h_X$ defined on 
the lattice $\Lambda$  with finite range interactions  supported in $X$.  Then for arbitrary observables $O_A$ and $O_B$ 
supported in the disjoint sets $A$ and $B$ we have
\begin{equation} \label{LR bound}
||[O_A(t),O_B(t)]||\leq c e^{-a(d(A,B)-v|t|)},
\end{equation}
where $d(A,B)$ is the distance between the sets $A$ and $B$ and $||O||$ is the operator norm of the observable. 
In the above equation $v$ is called Lieb-Robinson velocity and $a$ is just a constant.  As it is clear from the 
right hand side of the equation, for $d(A,B)>v|t|$ we expect an exponentially small left hand side which is a direct 
manifestation of cone-like horizon. For our models the Lieb-Robinson bound can be applied only for type-2 dispersion 
relation because this dispersion relation corresponds to exponentially decaying interactions. For the other cases 
we have a generalized Lieb-Robinson bound \cite{Generalized Lieb-robinson} that we now define: consider the Hamiltonian $H_{\Lambda}$ with interactions 
satisfying $\sum||h_X||\leq \lambda(1+d(r,r'))^{-(d+\alpha)}$ (for the power-law decaying interactions $\frac{1}{(d(r,r'))^{d+\alpha}}$ one 
can show that this condition is satisfied as far as $\alpha>0$ \cite{theory dynamic of entanglement}) then we have
\begin{equation} \label{GLR bound}
||[O_A(t),O_B(t)]||\leq c \frac{ e^{v|t|}-1}{(1+d(A,B))^{d+\alpha}}.
\end{equation}
As it is clear from the right hand side of the above equation one can not expect cone-like causal region in this case. However,
some other kinds of causal regions have been anticipated in the recent paper \cite{Eisert2013} for systems with long-range interactions.
As far as we take $\alpha>0$ for the massive and massless type-1 dispersion relation our model satisfies the assumptions of the 
theorem and so one can try to investigate the consequences of the generalized Lieb-Robinson bound. However, this is not 
the path that we are going to choose, instead we will show that most of the results can be understood with a simple semiclassical 
approximation which is just an intuitive way of understanding most of the results concerning the Lieb-Robinson bound.

\section{Correlations after a mass quench}
The quench protocol that we will consider is, as typically, the following: 
we initially prepare the system at the ground state of the Hamiltonian (\ref{Quadratic Hamiltonian}) 
with a dispersion relation $\omega_0$ and at $t=0$ we change suddenly the dispersion relation to 
$\omega$ (more specifically we will quench the mass parameter of the dispersion relation). The two point correlation function after such a general quench of (any parameter of) the dispersion relation is given by \cite{CardyCalabrese2007} 
\begin{align} \label{Two point function after quench}
& G(r,t)\equiv \langle\phi(r,t)\phi(0,t)\rangle - \langle\phi(r,0)\phi(0,0)\rangle =\nonumber\\
& \int \frac{d^d k}{(2\pi)^d} e^{i\mathbf{k}.\mathbf{r}}
\frac{(\omega_0^2(k)-\omega^2(k))(1-\cos(2\omega(k)t))}{\omega^2(k)\omega_0(k)}.
\end{align}
Initial correlations have been subtracted in the definition of $G(r,t)$, so that we can focus on the 
correlations that develop due to the quench. 
Details of the derivation of this formula can be found in \cite{CardyCalabrese2007}. 
Eq.~(\ref{Two point function after quench}) can be split into a time independent and a time dependent part
\begin{align}
\bar{G}(r) & =
\int \frac{d^d k}{(2\pi)^d} e^{i\mathbf{k}.\mathbf{r}}
\frac{(\omega_0^2(k)-\omega^2(k))}{\omega^2(k)\omega_0(k)}, \\
\tilde{G}(r,t) & =
-\int \frac{d^d k}{(2\pi)^d} e^{i\mathbf{k}.\mathbf{r}}
\frac{(\omega_0^2(k)-\omega^2(k))}{\omega^2(k)\omega_0(k)} \cos(2\omega(k)t).
\end{align}
As long as the integral in $\bar{G}(r)$ is convergent, the integral in $\tilde{G}(r,t)$ vanishes for large $t$, by virtue of the 
lemma Riemann-Lebesgue, and therefore ${G}(r,t)$ becomes stationary. If instead $\bar{G}(r)$ is not convergent, then ${G}(r,t)$ may
not become stationary for large times, i.e. it may exhibit persistent oscillations or increase indefinitely with time. 

We are interested in studying the consequences of a mass 
quench in the separate cases of the two dispersion relations introduced above. 
Before we proceed to the derivation of our results, we can anticipate some of them based on a semiclassical approach introduced next.

\subsection{Semiclassical approximation}\label{sec.semiclassical}

\begin{figure}[ht]
\begin{center}
\includegraphics[width=0.9\linewidth]{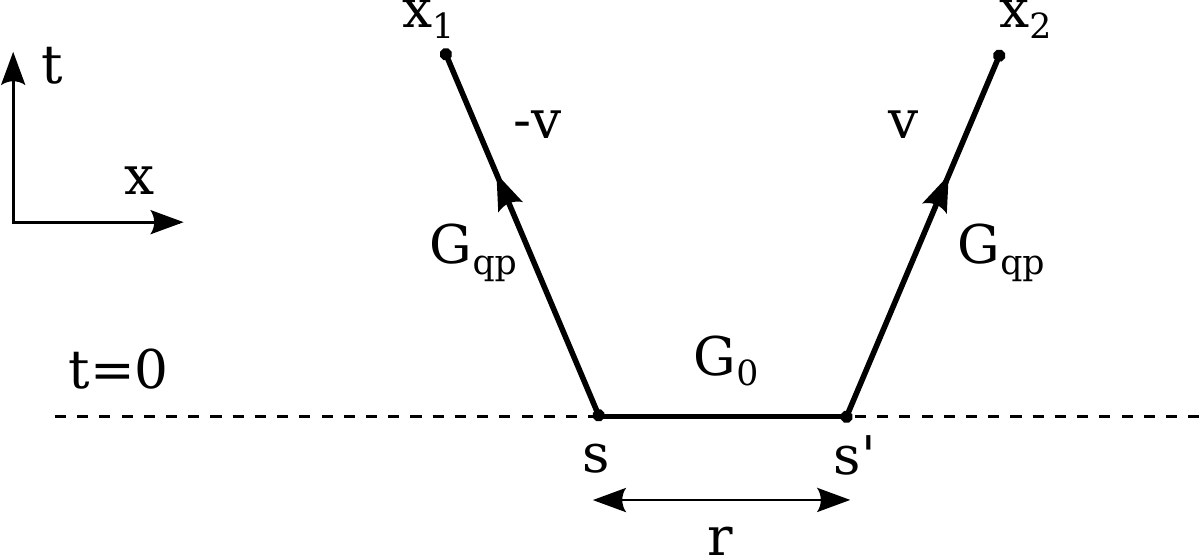}
\caption{\label{fig.semiclassical}Physical interpretation of the evolution of correlations after a quantum quench in the semiclassical 
approximation.}
\end{center}
\end{figure}

The semiclassical approximation provides a general physical interpretation of several features of quantum quenches, not only 
qualitatively but also quantitatively \cite{CardyCalabrese2007}. 
This has been already applied to the case of coupled harmonic oscillators, 
conformal field theories, spin chains and some interacting integrable models \cite{CC06,CC05,RI11,FC08,Ev13}. A quantum 
quench in a system where both the pre-quench and the post-quench Hamiltonians are 
essentially non-interacting (free or interacting but mappable to free) and invariant under spatial translations and under parity of
the fields, has as a result the instantaneous production of pairs of quasiparticles with opposite 
momenta (this is because under the above conditions the relation between the pre-quench and post-quench creation and annihilation
operators is a simple Bogoliubov transformation). 
If the initial state has short-range correlations then both quasiparticles in each of these pairs 
are produced within a distance of the order of the initial correlation length $\xi$.  
If instead the initial state has long-range correlations then quasiparticle pairs can be emitted from points at any distance. 
After the quench, the two correlated quasiparticles of each pair spread the initial correlations by traveling through the system. 
According to the semiclassical approximation, the quasiparticles move ballistically 
with velocity given by the group velocity $v(k) = d\omega(k)/dk$ corresponding to their wavelength $k$. This physical mechanism is 
pictorially shown in Fig.~\ref{fig.semiclassical}. 

The evolution after the quench can be explained by the collective effect of the quasiparticle motion. For example, let us assume that 
the initial correlations are short-range with a vanishingly small correlation length $\xi\to 0$ (which means that there are zero 
initial correlations between any two different points) and that after the quench the dispersion relation $\omega(k)$ has a maximum group 
velocity $v_{max}$, as in the relativistic case. Then the equal time two-point correlation function is expected to exhibit the horizon effect:
correlations between two points at distance $r$ are frozen to their initial zero value until the time $t^*(r)=|r|/(2v_{max})$ when a pair 
of the fastest quasiparticles emitted from the middle between these two points, reach them and transmit the first signal of the quench. 
If $\xi$ is not small, the first quasiparticle pairs to reach the two points, would be emitted not from the same but from points 
at distances of the order of $\xi$ and the correlations out of the 
horizon, i.e. in the region $t<t^*(r)$, will not be zero but will decay exponentially with the distance, as in the Lieb-Robinson theorem. 
If the initial correlations are long-range, i.e. $\xi\to\infty$ and they decay as a power law, one would expect that correlations in the
region $t<t^*(r)$ decay also as a power law with the distance. In this case where the correlations behave differently inside and outside
of the horizon, but do \emph{not} decay exponentially outside, the Lieb-Robinson bound is not present and the horizon is ``fake''. 
Finally, if the post-quench dispersion relation does \emph{not} have a maximum group velocity, then correlations at distant points develop 
immediately after the quench and there is no horizon at all. This is a physical explanation of the horizon effect.

More specifically, according to the above picture a semiclassical expression for the correlation function is of the form 
(Fig.~\ref{fig.semiclassical})
\begin{align}
G_{sc}(x_1,x_2;t) & = \int ds ds' \, G_{qp}(x_1,s;t) G_0(s,s') G_{qp}(x_2,s';t)  \nonumber \\
& = \int dr \, G_{qp}((x_2-x_1-r)/2;t)^2 G_0(r) \label{semicl} 
\end{align}
where $G_{qp}(x,x';t)$ is the semiclassical propagator of a quasiparticle moving from $x'$ to $x$ over time $t$ 
after the quench with velocity $v=|x'-x|/t$ and $G_0(x,x')$ are the initial correlations between points $x$ 
and $x'$. The integration finally runs over only one coordinate variable $r\equiv|s'-s|$ since the quasiparticle
 pairs have opposite velocities $v$ and $-v$, constraining geometrically the coordinates $s$ and $s'$. Obviously, 
by translational invariance each of the functions $G_{sc},G_{qp}$ and $G_0$ depend only on the coordinate distances, apart 
from the time. Moreover since the propagation of quasiparticles is ballistic, $G_{qp}(x,x';t)$ depends only on the 
velocity $v=|x'-x|/t$ and, if the dispersion relation has a maximum group velocity $v_{max}$, $G_{qp}(x,x';t)$ vanishes 
for $|x'-x|>v_{max}t$. In this case from (\ref{semicl}) we see that, in the region $|x_2-x_1|>2v_{max}t$, $G_{sc}(x_1,x_2;t)$ 
decays at large distances in the same way that $G_0(x_1,x_2)$ decays at large distances. In particular, if the latter decays exponentially 
the same holds for $G_{sc}(x_1,x_2;t)$ and therefore it exhibits a horizon in the strict sense. If instead $G_0(x_1,x_2)$ decays as 
a power law, then $G_{sc}(x_1,x_2;t)$ decays also as a power law and the horizon is fake. In all other cases, no horizon effect should be 
expected.

\begin{figure}
\begin{center}
\includegraphics[width=0.9\linewidth]{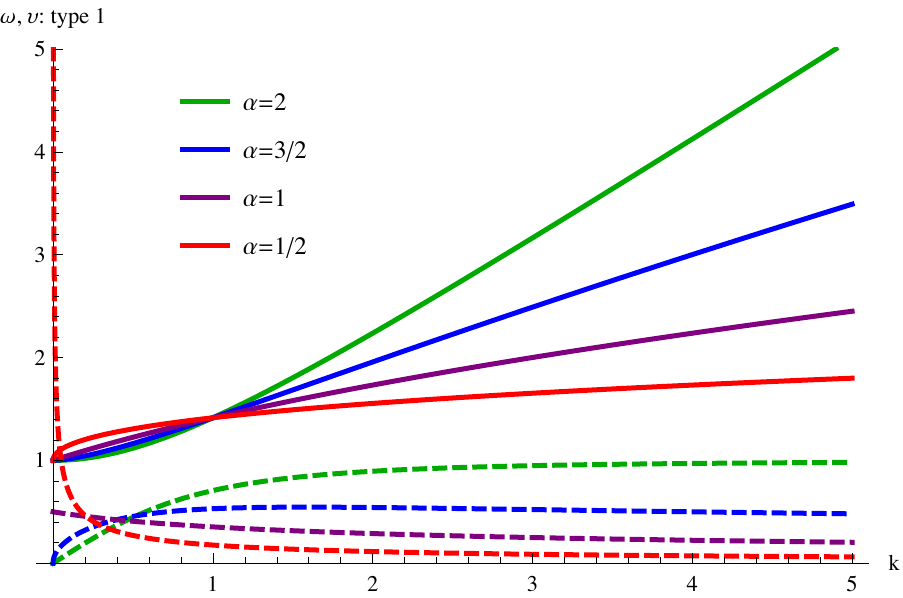}
\includegraphics[width=0.9\linewidth]{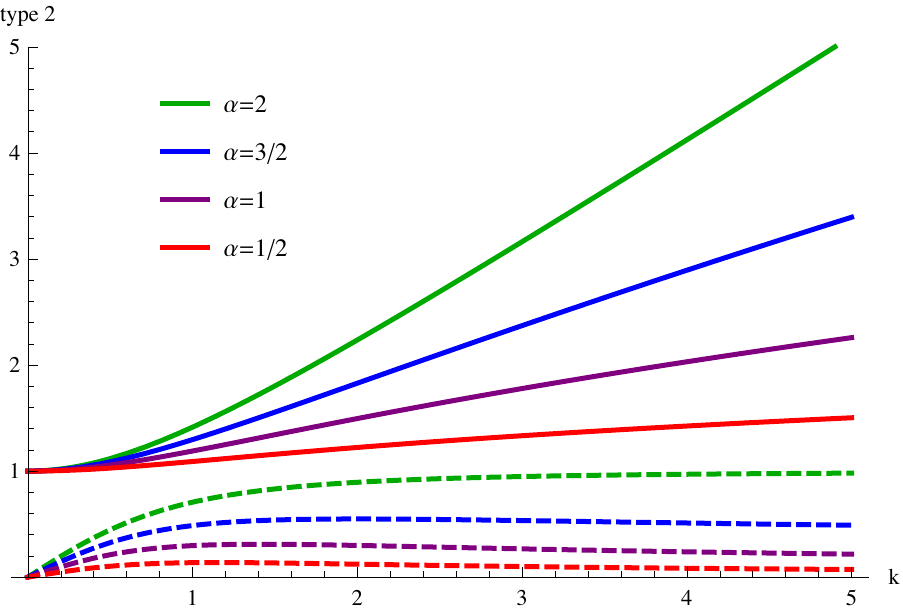}
\includegraphics[width=0.9\linewidth]{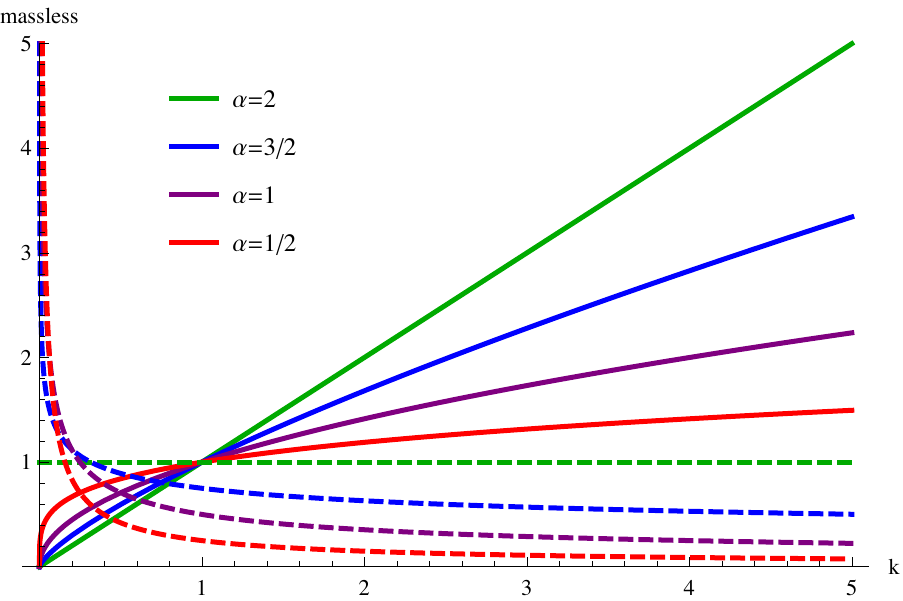}
\caption{\label{fig.SC}(Color online) Typical plots of the two types of dispersion relations $\omega_1(k), \omega_2(k)$ and the massless case,
along with the corresponding group velocity $v(k)=d\omega/dk$ as functions of $k$ for several values of the parameter
 $\alpha$ (2, 3/2, 1 and 1/2). The solid lines are plots of $\omega(k)$ and the dashed lines are plots of $v(k)$. The mass parameter 
 is set equal to 1 (unless zero).}
\end{center}
\end{figure}

In the present case it is simple to see that the group velocity for the above two types of dispersion relations has the following form 
(Fig.~\ref{fig.SC})
\begin{eqnarray} \label{Group velocity1}
v_1(k)&=&\frac{\alpha|k|^{\alpha-1}}{2\sqrt{|k|^{\alpha}+m^\alpha}}\\
\label{Group velocity2}
v_2(k)&=&\frac{1}{2}\alpha |k|(k^2+m^2)^{\frac{\alpha}{4}-1}.
\end{eqnarray}

From these expressions we deduce that in the massless case there is no maximum group velocity for $0<\alpha<2$ but instead the group 
velocity diverges as $k\to0$. In the massive case, for type-1 dispersion relation, there exists a maximum group velocity only for 
$\alpha\geq 1$ while for $\alpha<1$ it also diverges as $k\to0$. For type-2 there always exists a maximum group velocity. 
This means that, according to the discussion above, we should expect that in the case of a type-2 massive-to-massive quench, 
causality holds and the horizon effect is present, since both conditions are satisfied: the initial correlations are exponentially 
decaying and the evolution is governed by an upper-bounded group velocity of excitations. In the case of a type-1 massive-to-massive 
quench the initial correlations decay as a power law and for $\alpha\geq 1$ there is also a maximum group velocity in this type of 
dispersion relation, so we expect that there is a fake horizon. In all other cases, no horizon effect should be expected.

Apart from the explanation of the horizon effect, the semiclassical approximation explains also the equilibration at large times. Indeed, 
if $\omega(k)$ has a local minimum or, more generally, a stationary point, then the destructive interference (\emph{dephasing}) of all 
incoherent quasiparticles that finally arrive at any point from very distant initial points, has as a result the stationary behavior of 
the correlations at large times, according to a simple stationary phase argument. In the present problem, $\omega(k)$ has a stationary point 
(more specifically a minimum at $k=0$) only in the massive case of type-2 dispersion relation, therefore equilibration is expected to occur 
in this case. In the massless dispersion relation or the massive type-1, the minimum at $k=0$ is not smooth but non-analytic, however 
application of the stationary phase method is still possible using a change of integration variable, so equilibration may occur also in 
this case. In fact, as we will see below, equilibration 
occurs in all massive-to-massive quenches in both types of dispersion relations, as well as in the massless case for certain ranges of 
values of $\alpha$.

\subsection{Exact results}

We now proceed to the study of the behavior of $G(r,t)$ given by (\ref{Two point function after quench}), using both analytical 
and numerical methods. 
There are two distinct cases to study: the massless case $m=0$ and the massive one $m>0$, where $m$ is the post-quench mass parameter. 
These should be studied separately as they exhibit different convergence properties. In addition, we should distinguish between the two 
different types of dispersion relations. Firstly, we will consider the case where the dispersion relation is of type-1, in which the 
initial correlations are long-range, i.e. decay as a power law at large distance. Later, in subsection \ref{The effect of the initial state} 
we will consider the case of type-2 dispersion relation instead, thus studying also the effect of the initial state on the dynamics.

\subsubsection{Massless post-quench Hamiltonian}
\label{Massless post-quench Hamiltonian}

\begin{figure}[h!]
\begin{center}
\includegraphics[clip,width=0.9\linewidth]{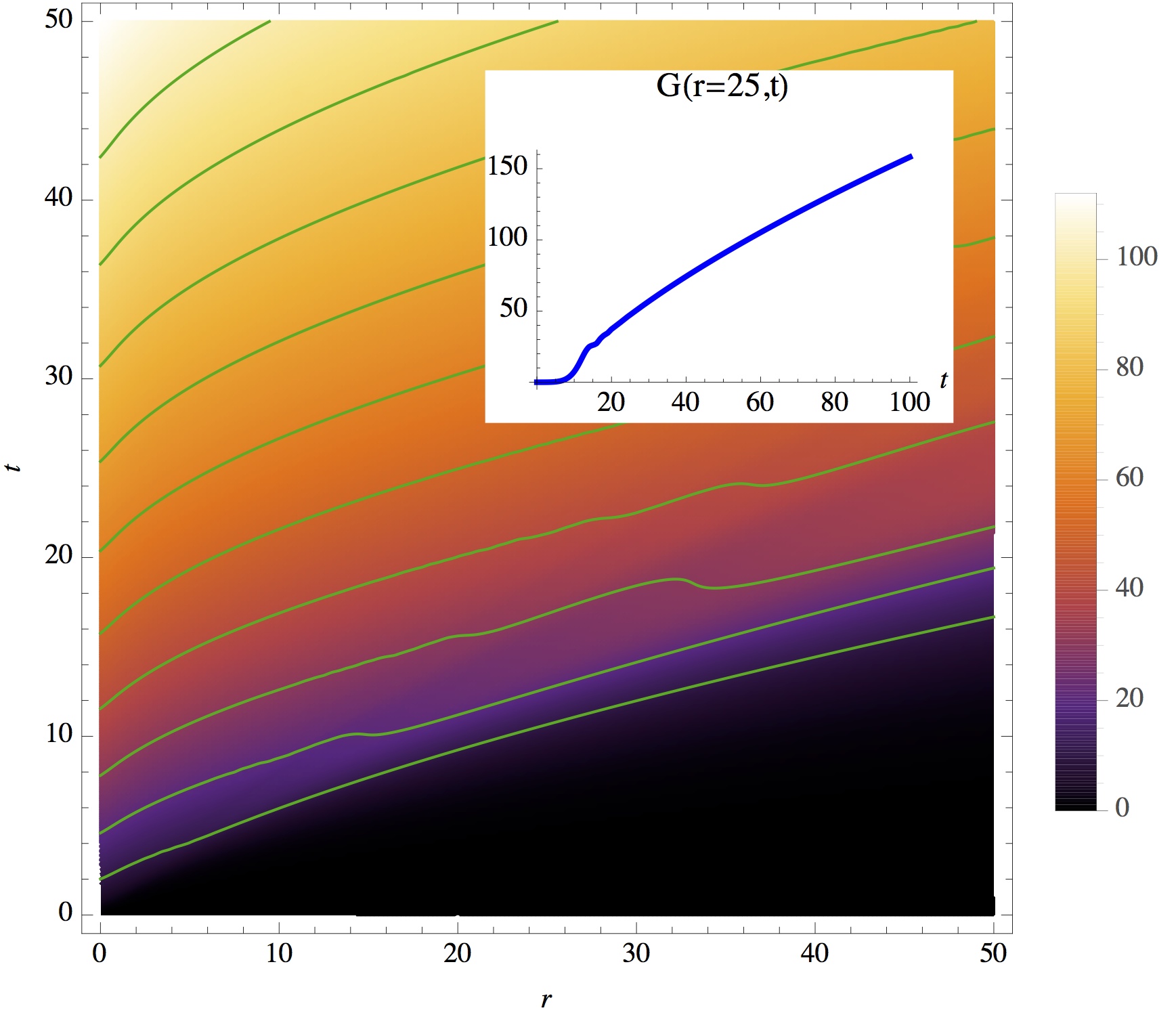}\\
\includegraphics[clip,width=0.9\linewidth]{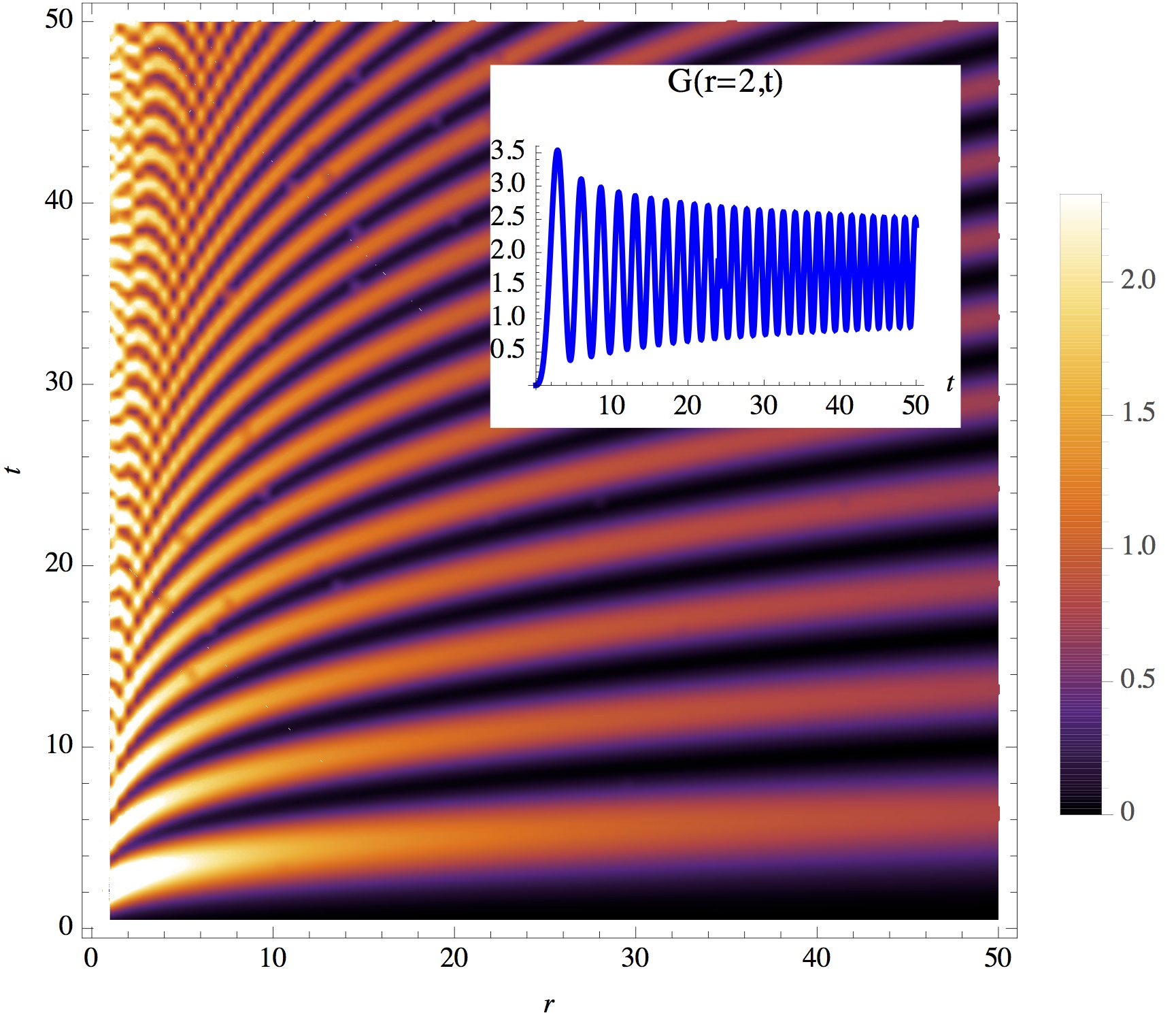}\\
\caption{\label{fig.1a} (Color online) Dependence of the evolution of correlations on the parameter $\alpha$. 
The plots are density plots of the correlations ($|G(r,t)|$ as given by (\ref{Two point function after quench})) in space-time, for the quench from $m_0=1$ to zero mass in type-1 dispersion relation and $1d$. 
The parameter $\alpha$ takes values $\frac{3}{2}$ (top) and $\frac{1}{2}$ (bottom). 
In the upper plot contours are also drawn. The insets show plots of the correlation function as a function of $t$ at 
some fixed distances ($r=25$ and $r=2$ respectively). The correlations increase with time for $\alpha=\frac{3}{2}$, 
while for $\alpha=\frac{1}{2}$ they do not. For $\alpha=\frac{1}{2}$ the system equilibrates, even though as we see in 
the inset, the time decay of the oscillations is very slow. Note that since the results are symmetric with respect to the vertical
axis we plotted just the positive side of the real line.}
\end{center}
\end{figure}

\begin{figure}[ht!]
\begin{center}
\includegraphics[clip,width=0.9\linewidth]{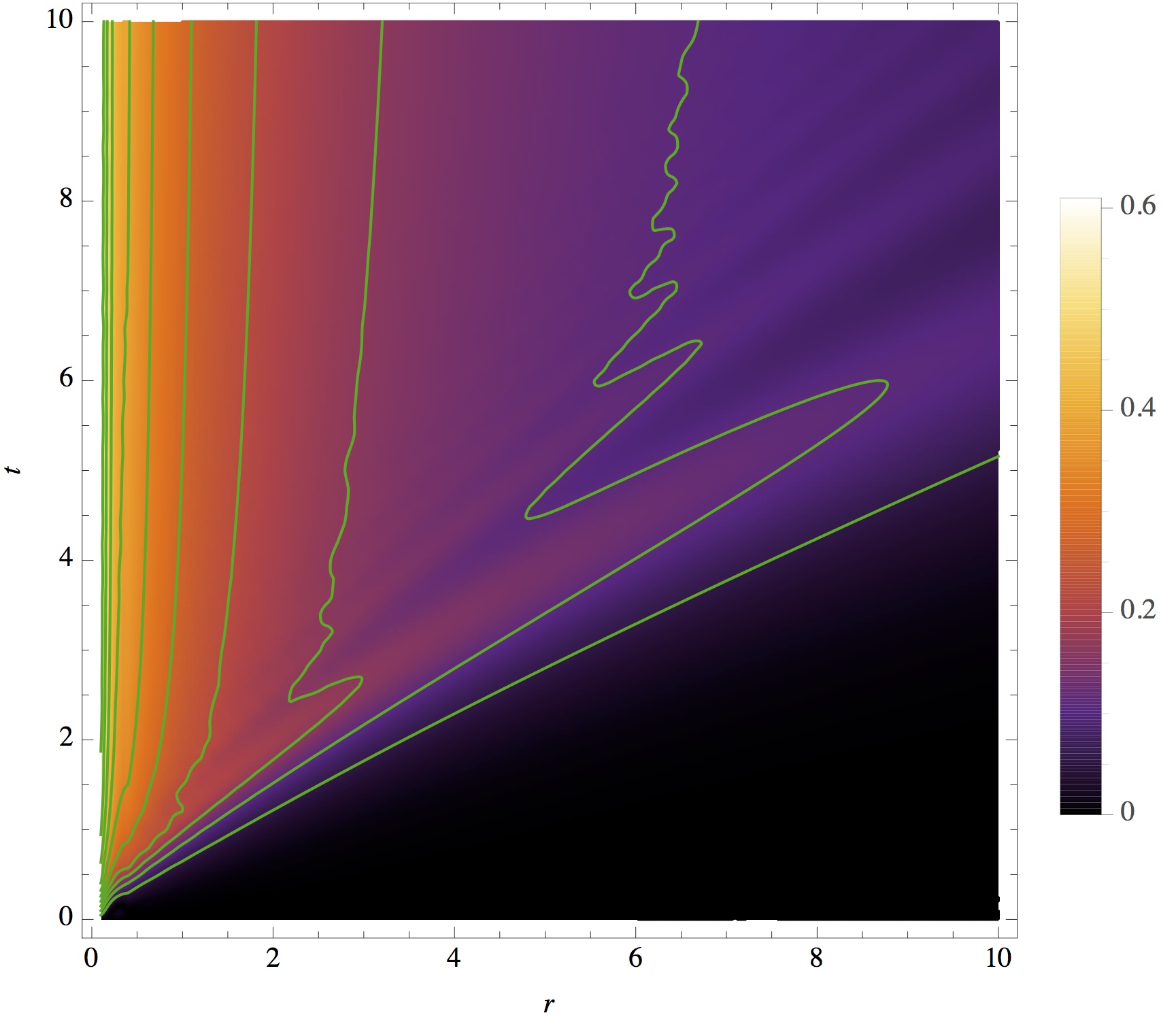}\\
\includegraphics[clip,width=0.9\linewidth]{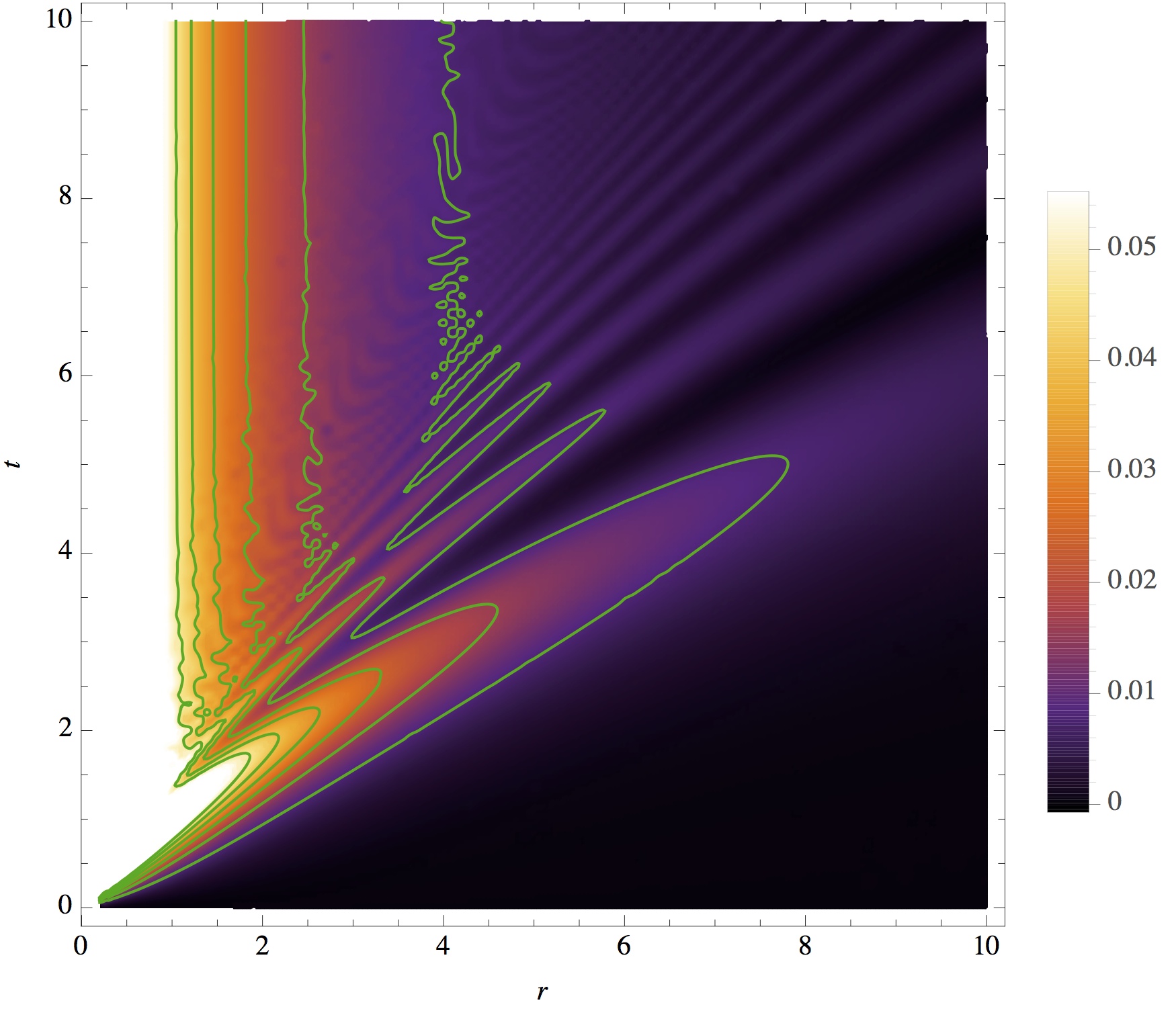}\\
\caption{\label{fig.1b} (Color online) The effect of dimensionality on the evolution of correlations. \\
The plots are density and contour plots of the correlations in space-time for the quench to zero mass in type-1 dispersion relation 
with $\alpha=\frac{3}{2}$ and $m_0=1$ in 
$2d$ (top) and $3d$ (bottom). The contours become asymptotically vertical straight lines for large times, 
meaning that the correlations tend to stationary values. Compare with Fig.~\ref{fig.1a}.}
\end{center}
\end{figure}

Let us focus on the massless case $m=0$, assuming that before the quench the system is described by the type-1 dispersion relation.
The integral in $\bar{G}(r)$ is written as
\begin{align} \label{Two point function after deep quench zero final mass}
\bar G(r) =
\int \frac{d^d k}{(2\pi)^d} e^{i\mathbf{k}.\mathbf{r}}\frac{m_0^{\alpha}}{|k|^{\alpha}(m_0^{\alpha}+|k|^{\alpha})^{\frac{1}{2}}}.
\end{align}
and it is convergent for all values $d-1\leq\alpha<d$. For $\alpha<d-1$ it has an ultraviolet (UV) divergence, which however can be 
ignored since it is not relevant for the macroscopic behavior of the system. By introducing a suitable UV cutoff we can regularize 
the integral without spoiling its (physically interesting) macroscopic behavior. On the other hand, for $\alpha>d$ the integral 
exhibits an infrared (IR) divergence. In one spatial dimension ($d=1$) this means that there is equilibration for $\alpha<1$, 
while for $\alpha>1$ we find that ${G}(r,t)$ does not become stationary but increases indefinitely with time instead, as it can be easily 
seen also numerically. In two spatial dimensions ($d=2$) the above constraint means that there is always equilibration for $\alpha<2$ 
while for the short-range case $\alpha=2$ it is known that $G(r,t)$ increases logarithmically with time \cite{CS2010}. Lastly, in three spatial 
dimensions ($d=3$), $G(r,t)$ equilibrates for any value of $\alpha$ (as already explained, the case $\alpha>2$ is not physically interesting 
in any dimension).

In the cases where equilibration occurs, the approach of the correlation function $G(r,t)$ to its stationary large-time value $\bar G(r)$, 
is given by the decay of the time dependent part $\tilde G(r,t)$ for large times $t$, with the distance $r$ kept fixed. The asymptotic 
behavior of this integral is 
\be
\tilde G(r,t \to \infty) \sim \frac{1}{t^{2({d}/{\alpha})-2}} \label{G10}
\ee
i.e. a power law with an exponent dependent both on $d$ and $\alpha$ (the asymptotic behavior of such integrals, as well as 
of the ones below, can be derived by suitable choice of the integration variable and application of the stationary phase method 
or the steepest descent method which in most cases reduces to a simple Wick rotation of the integration variable in the complex plane). 
On the other hand, the large-distance behavior of the stationary correlations $\bar G(r)$ turns out to be
\be
\bar G(r \to \infty) \sim \frac{1}{r^{d-\alpha}}, \qquad \alpha<d. \label{G10s} 
\ee

We have checked the behavior of the system regarding equilibration and we will now check whether it exhibits a horizon. 
As mentioned in the introduction the Lieb-Robinson causal region is characterized by the exponential way in which the correlations 
decay outside the causal region. The presence of horizon can be checked by studying the behavior of $G(r,t)$ keeping $t$ fixed and taking 
the limit $r\to\infty$. We find that the decay at large distance is given by
\be
G(r \to \infty,t) \sim \frac{1}{r^{d+\alpha}}.  
\ee
Since this decay is not exponential, there is no horizon. Fig.~\ref{fig.1a} and~\ref{fig.1b} show typical plots of the correlations 
in 1, 2 and 3 spatial dimensions.

The above results are in agreement with the semiclassical approximation. Indeed, the absence of causality and the associated horizon
effect can be attributed to the fact that the group velocity is unbounded for all values of $\alpha$ in the massless case. On the other 
hand, since the dispersion relation has no stationary point but a non-analytic minimum at $k=0$, naively we would not expect equilibration 
at large times. However a simple change of variables allows us to apply the stationary phase method from which we find that equilibration 
\emph{does} occur.

\subsubsection{Massive post-quench Hamiltonian}

\begin{figure}[h!]
\begin{center}
\includegraphics[clip,width=0.9\linewidth]{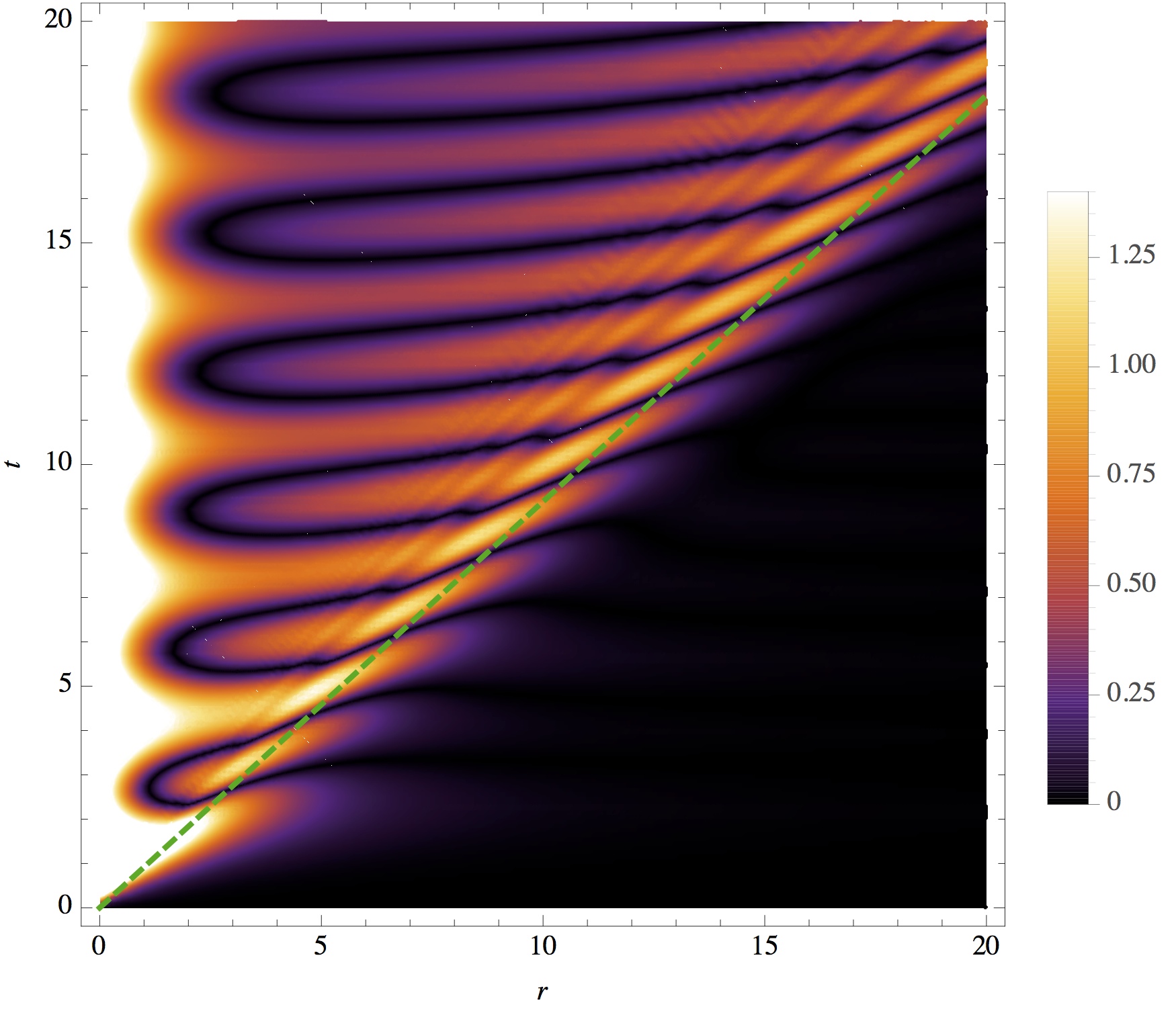}\\
\includegraphics[clip,width=0.9\linewidth]{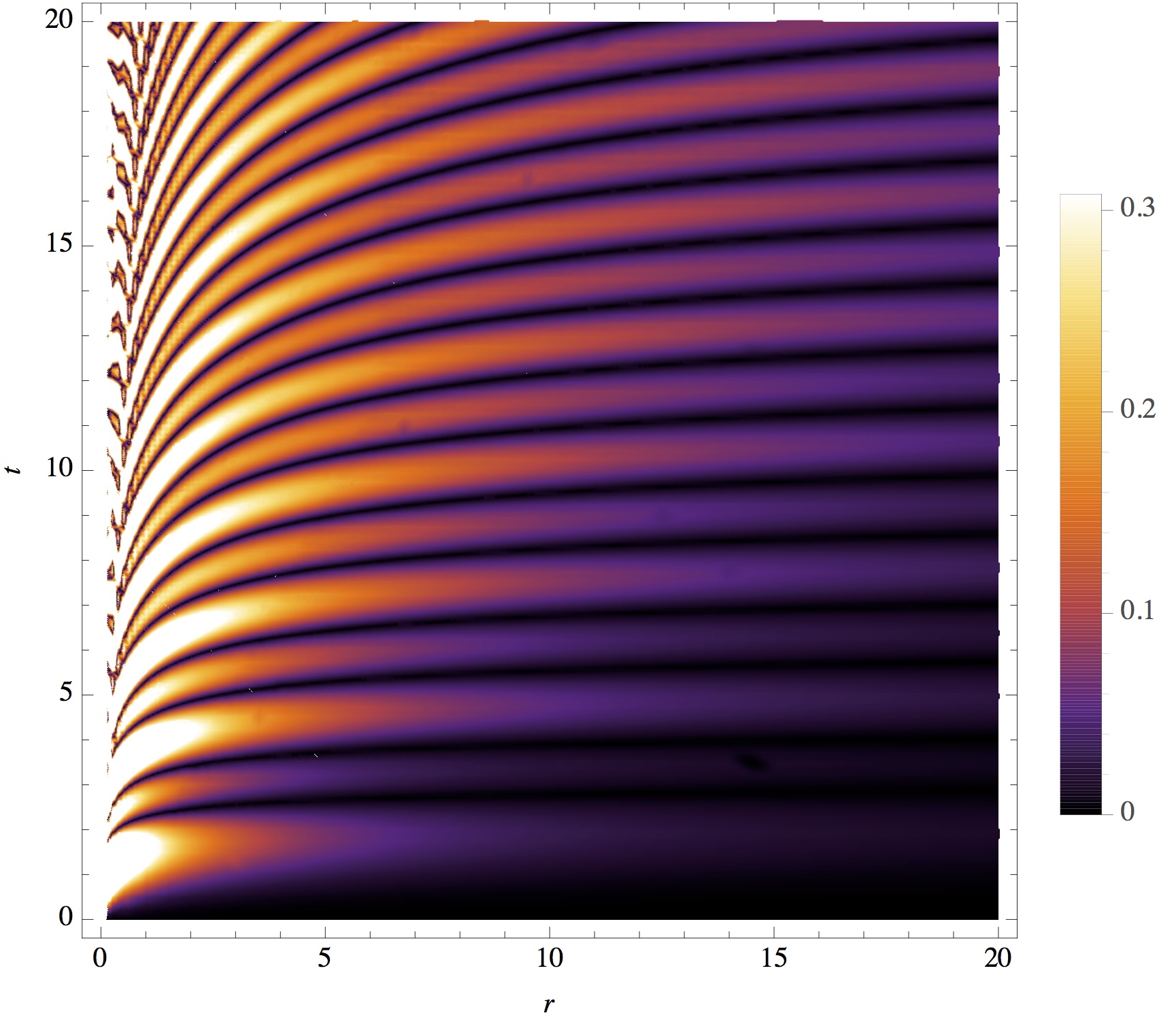}\\
\caption{\label{fig.massive} (Color online) Density plots of correlations for the quench from massive-to-massive in type-1 case in $1d$. 
The parameter $\alpha$ is $\frac{3}{2}$ (top) and $\alpha=\frac{1}{2}$ (bottom) with $m_0=2$ and $m=1$. 
\emph{Top:} The evolution seems to exhibit a horizon, however this is not true horizon since the decay of correlations 
with the distance turns out to be a power law. The dashed green line indicates the slope that corresponds to the maximum group 
velocity $v_{max}=\sqrt{3}/2/2^{2/3} \approx 0.5456$ for this particular value of $\alpha$, as derived from Eq.(\ref{Group velocity1}). 
\emph{Bottom:} There is no sign of horizon in this case. 
Even if this is not evident from the density plots due to the limited temporal range, the system equilibrates in both cases.}
\end{center}
\end{figure}

Let us switch to the massive case $m>0$. 
In this case there are no IR divergences in $\bar G(r)$ and so $\tilde G(r,t)$ decays always to zero for large $t$ and the system 
equilibrates. The small $k$ behavior of the  type-1 dispersion relations is (Fig.~\ref{fig.SC})
\begin{align}
 \omega_1(k) \approx m^{\alpha/2} + \frac12 \frac{k^\alpha}{m^{\alpha/2}} + ...
\end{align}
and the stationary phase method gives 
\begin{align} 
&\tilde G(r,t\to\infty) \sim {t^{-d/\alpha}} \cos(2m^{\alpha/2}t) , \label{G11} 
\end{align}
The large distance decay of the stationary correlations $\bar G(r)$ is given by a power law
\be
\bar G(r \to \infty) \sim \frac{1}{r^{d+\alpha}}. \label{G1s} 
\ee
due to the branch cut of the integrand in the complex $k$-plane, which starts from the origin $k=0$. The same asymptotic 
behavior is valid for large distances at any fixed time $G(r \to \infty,t)$.

As far as the horizon effect is concerned, we find that for $\alpha<1$ correlations start developing even at arbitrarily 
long distances soon after the quench, therefore no horizon exists, while for $\alpha>1$ they seem to form a horizon, however 
their decay outside of it is power law therefore even in this case no true Lieb-Robinson horizon occurs. This is demonstrated 
in Fig.~\ref{fig.massive} where we show plots for the values $\alpha=\frac{3}{2}$ and $\alpha=\frac{1}{2}$. As before, the 
semiclassical approximation can explain this behavior. According to the discussion of section \ref{sec.semiclassical}, since 
the initial correlations for the type-1 dispersion relation are long-range, i.e. decay as a power-law with the distance, the same 
is true for the evolution of correlations after the quench and a true horizon effect is not possible. However, since the group 
velocity has a maximum for $\alpha\geq1$, a generalized horizon with power-law decaying correlations outside of it is present in this case. 
In contrast, for $\alpha<1$ the group velocity is unbounded and there is no horizon at all.

\subsubsection{The effect of the initial state}\label{The effect of the initial state}

The same analysis can be repeated in the case where the dispersion relation is of type-2. 
According to (\ref{propagator omega2 assymptotic}) the initial correlations are then short-range, i.e. 
they decay exponentially with the distance.
\begin{figure}[h]
\begin{center}
\includegraphics[clip,width=0.9\linewidth]{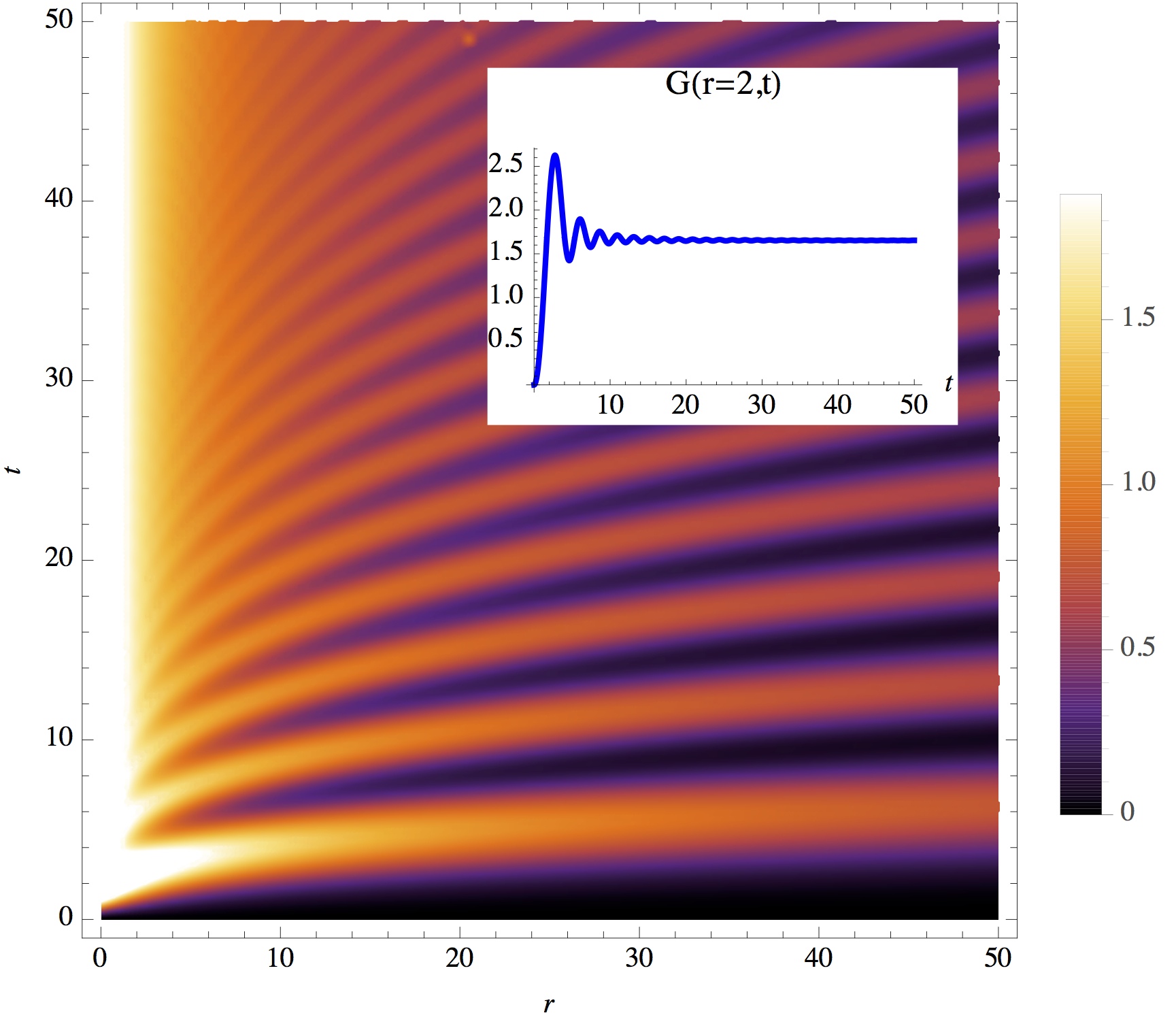}
\caption{\label{fig.massless type 2} (Color online) Density plot of correlations for the massive-to-massless quench in type-2 
dispersion relation with $\alpha=\frac{1}{2}$, $m_0=1$ in $1d$. There is no sign of horizon. The inset shows a plot of the correlation
function as a function of time $t$ for a fixed distance $r=2$. Compare with the bottom plot of Fig.~\ref{fig.1a} which corresponds to 
type-1. Notice that the decay of oscillations is much faster for type-2 than for type-1.}
\end{center}
\end{figure}

In the massive-to-massless case, as in the type-1 case of subsection \ref{Massless post-quench Hamiltonian}, 
equilibration occurs only for $\alpha<d$ since the integral in $\bar G(r)$ is then finite, while otherwise it is IR divergent. 
Furthermore, there is no horizon effect in this case either, as we would also expect following the semiclassical analysis. 
Fig.~\ref{fig.massless type 2} shows a typical plot of such a quench.

In the massive-to-massive case, again similarly to the type-1 case, equilibration occurs for all values of $\alpha$'s, 
since the integral in $\bar G(r)$ is always finite. The large time decay turns out to be 
\begin{align}
&\tilde G(r,t\to\infty) \sim {t^{-d/2} \cos(2m^{\alpha/2}t)}.
\end{align}
In addition the stationary correlations at large times are
\begin{align}
&\bar G(r\to\infty) \sim r^{\frac\alpha 4 - \frac{d+1}{2}} e^{-m r}, \quad m_0>m,
\end{align}
where in the last case we assumed (as is typically done) that $m_0>m$.

\begin{figure}
\begin{center}
\includegraphics[clip,width=0.9\linewidth]{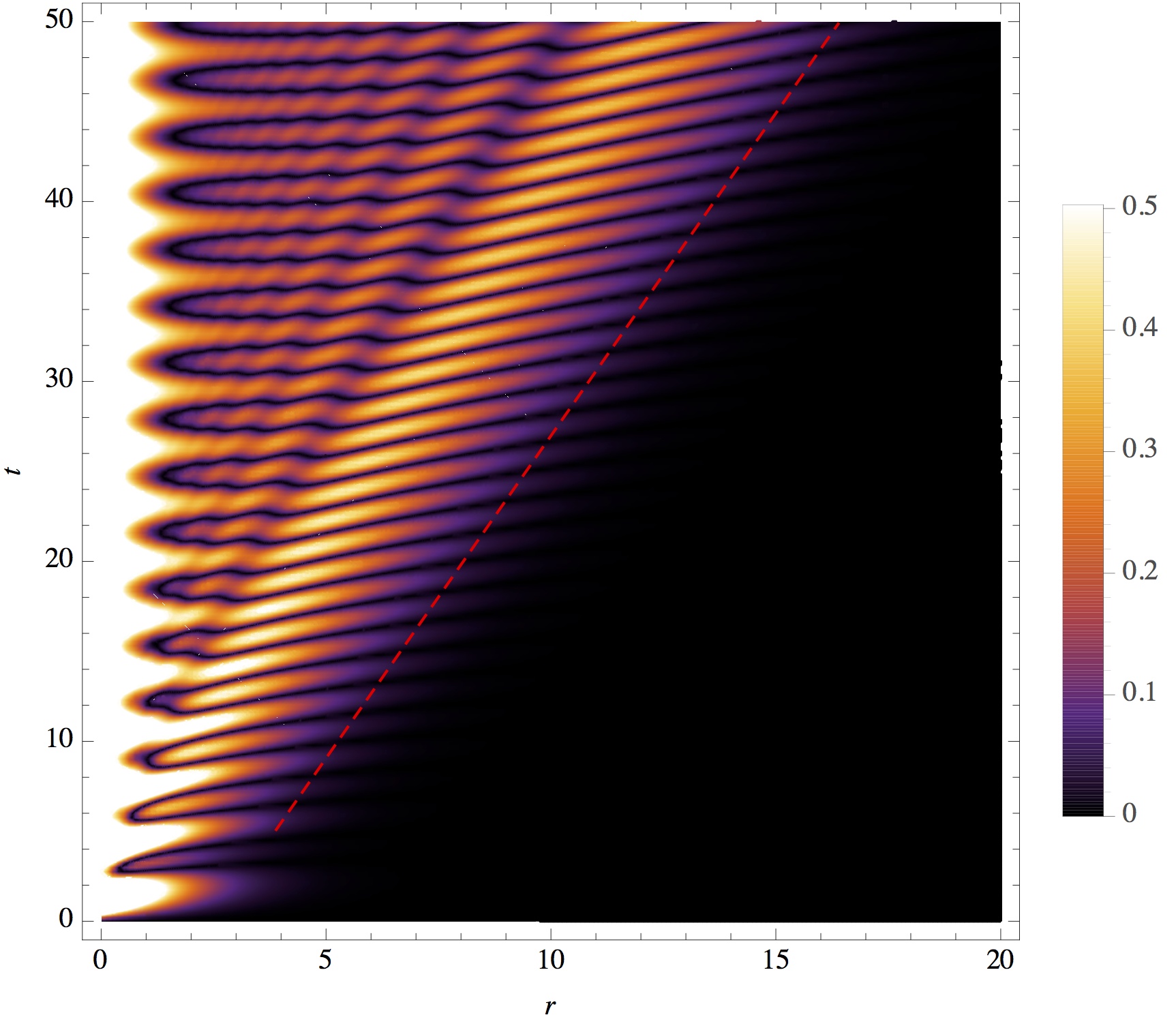}\\
\caption{\label{fig.2} (Color online) Density plots of correlations for the quench from massive-to-massive in type-2 dispersion relation. 
We used the parameter values $\alpha=\frac{1}{2}$, $m_0=2$ and $m=1$ in $1d$. The plot demonstrates the existence of causal region with 
linear horizon. 
The dashed red line is an estimate of the horizon based on contour lines. From the slope of 
this line we find that the maximum group velocity is approximately $v_{\text{max}}\approx 0.14$ which is in perfect 
agreement with the exact value $v_{\text{max}}= 3^{3/8}/7^{7/8}/2\approx 0.1375$ as derived from Eq.(\ref{Group velocity2}). 
}
\end{center}
\end{figure}

Lastly, in type-2 massive-to-massive quench the horizon effect is prominent (Fig.~\ref{fig.2}) as in the type-1 case.
However it is worth to mention that, in contrast to type-1, this is a true horizon effect i.e. with exponentially decaying 
correlations outside of it. 
The reason for this exponential decay is that (as in the case of the ground state correlations under the type-2 dispersion relation)
 the integrand in $G(r,t)$ exhibits, in the complex $k$-plane, branch cuts that start from $k=\pm im$ and $k=\pm im_0$ and since $m_0>m$ 
the closest to the origin are the former. Once again, this behavior can be explained by the semiclassical approximation. 
Indeed, in massive type-2 dispersion relation there always exists a maximum group velocity and, in addition, the initial 
correlations decay exponentially with the distance, which according to the semiclassical approximation are the conditions 
for a true Lieb-Robinson horizon. This is demonstrated in Fig.~\ref{fig.2}, where the slope of the horizon is accurately 
predicted from the maximum group velocity of the post-quench dispersion relation.

\section{Long time limit and GGE}
\subsection{Validity of the GGE}

Having shown that, under specified conditions, the long-range systems we consider equilibrate for large times, we will now check whether 
the asymptotic stationary behavior can be described by the GGE. The GGE density matrix can be written as 
\begin{eqnarray} \label{GGE}
\rho_{\text{GGE}}=Z^{-1} {e^{-\sum \lambda_k I_k}}, \qquad Z=\text{Tr} \; e^{-\sum \lambda_kI_k},
\end{eqnarray}
where the $I_k$'s are a set of commuting and independent integrals of motions and the $\lambda_k$'s are the associated Lagrange 
multipliers, fixed by the self-consistency requirement that the values of $I_k$ are the same in the initial state and in the ensemble
\be
\langle I_k \rangle_{t=0} = \langle I_k \rangle_{\text{GGE}}.
\ee

A natural choice of integrals of motions that one can consider in the present problem is the number of particles with a 
certain momentum, i.e. $n_k=a_k^{\dagger}a_k$, where
\begin{eqnarray} \label{creation anihilation}
a^{\dagger}_k&=&\frac{1}{\sqrt{2\omega(k)}}(\omega(k)\phi_{-k}-i\pi_{-k}),\nonumber\\
a_k&=&\frac{1}{\sqrt{2\omega(k)}}(\omega(k)\phi_k+i\pi_k).
\end{eqnarray}
are the usual creation and annihilation operators. 
With this choice of integrals of motion, it is easy to see that the large-$t$ stationary part of the 
correlations $G(r,t \to \infty) = \bar G(r)$ is automatically equal to the GGE prediction \cite{SC14}. 
Indeed $\bar G(r)$ turns out to be directly related to the initial momentum density $\langle n_k\rangle_{t=0}$: 
it is essentially its Fourier transform. On the other hand, the GGE prediction for the two-point correlator is the 
Fourier transform of the momentum density in the GGE $\langle n_k\rangle_{\text{GGE}}$. Therefore by the very 
definition of the GGE that requires $\langle n_k\rangle_{GGE} = \langle n_k\rangle_0$, we see that, as long as the 
correlation function equilibrates (i.e. $\tilde G(r,t\to \infty) \to 0$) its stationary value $\bar G(r)$ is automatically 
equal to the GGE prediction.

The explicit values of the Lagrange multipliers $\lambda(k)$ are given by \cite{CardyCalabrese2007}
\be \label{Lagr_mult}
\lambda(k) = 2 \log\left|\frac{\omega_0(k)+\omega(k)}{\omega_0(k)-\omega(k)}\right|.
\ee

\subsection{Non-locality of the GGE}

We have just shown that the large time behavior of the 2-point correlation function is described by the GGE (\ref{GGE}) where the Lagrange multipliers are given by (\ref{Lagr_mult}).

The general belief is that the integrals of motion used in the GGE should be \emph{local} conserved charges, meaning integrals 
over the whole coordinate space of local density operators (see \cite{Fagotti2013} and references therein). In fact all quantum 
systems, even non-integrable ones, possess infinite integrals of motion: the projections onto their energy eigenstates, which, 
ignoring the case of degeneracies, contain all information about the initial state. An ensemble based on such projections onto 
all post-quench eigenstates is called the \emph{diagonal} ensemble and therefore describes correctly (but in a trivial way) 
the time average of any local observable (or its large time value, provided it equilibrates). However, a statistical ensemble 
is utile only if it is economic, in the sense that it does not require \emph{all} information about the initial state. Integrable 
models are special in that they possess a set of \emph{local} integrals of motion, which are much less in number than the dimension of 
the Hilbert space (their number grows only linearly with the system size) and are sufficient for their exact solvability. Therefore it 
is sensible to assume that they are also sufficient for the description of their large time behavior. Of course the above argument 
works equally well for any system with a \emph{minimal} set of integrals of motion that is smaller in number than the dimension of the 
Hilbert space but sufficient for its exact solvability, even if they are \emph{not} local. The locality requirement is usually 
introduced so that the GGE reduced density matrix corresponding to any subsystem can be constructed out of extensive quantities 
that therefore depend only on the size of that subsystem. This is because in standard thermodynamics we would expect the ``generalized free energy'' $F_{\text{GGE}}=-\log Z[\rho_{\text{GGE}}] $ to be extensive on the subsystem's size. (In fact this is a stronger requirement 
than the condition that $F_{\text{GGE}}$ can be written as a sum over all local integrals of 
motion, since such an infinite sum of local quantities is not necessarily local or extensive itself.)

However, if the Hamiltonian of the system is itself non-local, then it may not be possible to find a set of local integrals
of motion to construct the GGE. In such a case, according to the above discussion, if a GGE based on a minimal set of non-local
integrals of motion (less in number than the dimension of the Hilbert space) describes correctly the system's stationary behavior, 
it is also an economic ensemble, because it depends only on \emph{partial} information about the initial state. For example, 
it was recently pointed out that this is the case in the harmonic Calogero model \cite{RS14}. 

In the form (\ref{GGE}) the GGE does not seem to satisfy the locality requirement, since the conserved quantities from which it is constructed are the momentum occupation number operators $n(k)$ which are not local but global operators. However it may still be possible to rewrite them in a local form. One can easily see that in the earlier studied short-range case $\alpha=2$, the generalized free energy
$F_{\text{GGE}}=\int dk \; \lambda(k) n(k) $ can be cast as an infinite sum of local quantities. To see this, 
it is sufficient to write $n(k)$ in terms of local fields using its definition 
\begin{align}
n(&k) = a^\dagger_k a_k = \nonumber \\  
& \frac{1}{2}\left(\omega(k) {\phi}_{-k}{\phi}_{k} 
+i {\phi}_{-k}{\pi}_{k} - i {\pi}_{-k}{\phi}_{k} + 
{\pi}_{-k}{\pi}_{k}/\omega(k)\right) = \nonumber  \\
& \frac{1}{2} \int dx dy \; e^{ik(x-y)} [\omega(k) \phi(x) \phi(y) + \nonumber \\
& + i (\phi(x) \pi(y) - \pi(x) \phi(y)) + \pi(x) \pi(y) /\omega(k) ]
\end{align}
and substitute in the generalized free energy
\begin{align}
& F_{\text{GGE}} = \frac{1}{2} \int dx dy  \;  [ D_1(x-y) \phi(x) \phi(y) + \nonumber \\
& i D_2(x-y) \left( \phi(x) \pi(y) - \pi(x) \phi(y) \right)+ 
D_3(x-y) \pi(x) \pi(y) ] ,	\label{gfe}
\end{align}
where 
\begin{align}
D_1(s) & = \int dk \; e^{iks} \lambda(k) \omega(k), \nonumber \\
D_2(s) & = \int dk \; e^{iks} \lambda(k), \\
D_3(s) & = \int dk \; e^{iks} \lambda(k) / \omega(k) . \nonumber
\end{align}
Using the commutation relation $[\phi(x),\pi(y)] = i \delta(x-y)$ and the fact that $D_2(s)$ is an even function (since $\lambda(k)$ is such too) we can further write (\ref{gfe}) as 
\begin{align}
 F_{\text{GGE}} =  & \frac{1}{2} \int dx dy  \;  [ D_1(x-y) \phi(x) \phi(y) + \nonumber \\
& D_3(x-y) \pi(x) \pi(y) ] - \pi \lambda(0)
\end{align}
The last expression is a double instead of a single spatial integral. However, if both $\omega(k)$ and
$\lambda(k)$ can be Taylor expanded in terms of $k$ at $k=0$, then each of $D_1(s), D_2(s)$ and $D_3(s)$ can be expressed by means of
differential operators so that the whole expression is a single spatial integral. In fact it is sufficient that 
$\omega(k)$ and $\omega_0(k)$ are Taylor expandable because $\lambda(k)$, being a function of these two, is Taylor expandable whenever 
each of them is such. For example, in this case $D_1(s)$ reads
\begin{align}
D_1(s) & = \sum_{n=0}^\infty c_n \int dk \; k^n  e^{iks} \nonumber \\
& = \left[ \sum_{n=0}^\infty c_n \left( -i \frac{d}{ds} \right)^n \right] \; \int dk \; e^{iks}  \nonumber \\
& = 2\pi \; \mathcal{\hat{D}}^{(1)}_{s} \; \delta(s) ,
\end{align}
where 
\be
\mathcal{\hat{D}}^{(1)}_{s} \equiv  \sum_{n=0}^\infty c_n \left( -i \frac{d}{dx} \right)^n 
\ee
and 
\be
c_n = \frac{1}{n!} \frac{d^{n}}{dk^{n}} \big(\lambda(k) \omega(k)\big)\Big|_{k=0}.
\ee
Similarly $D_3(s)$ can be written as $D_3(s) = 2\pi \mathcal{\hat{D}}^{(3)}_{s} \delta(s)$, so that $F_{\text{GGE}}$ finally takes the form 
\be
F_{\text{GGE}} = \frac{1}{2} \int dx \;  \left[ \phi(x) \mathcal{\hat{D}}^{(1)}_{x} \phi(x) 
+ \pi(x) \mathcal{\hat{D}}^{(3)}_{x} \pi(x) \right] - \pi \lambda(0)
\ee
meaning that in the short-range case the generalized free energy $F_{\text{GGE}}$ is actually a single spatial integral of (an infinite sum of) local operators. 

Specializing now the above discussion in the case of long-range models that we study, we see that the massive type-1 
dispersion relation, $\omega_1(k)$ is Taylor expandable \emph{only} for $\alpha=2$, while the massive type-2 dispersion 
relation $\omega_2(k)$ is such for \emph{all} values of $\alpha$. Therefore in type-1 quench with both $m,m_0>0$, the 
GGE satisfies the above locality property only in the short-range case $\alpha=2$, while in type-2 with the same conditions 
it always satisfies that property. We arrive to the conclusion that,  for the model 
studied, the GGE is valid even in the cases that it cannot be expressed in terms of local charges. It is worth 
mentioning that the importance of considering non-local integrals of motion has been already emphasized in the study of  Cooper pair distribution
function after quantum  quench in fermionic superfluid problem \cite{Gurarie}, see also \cite{Levitov,Yuzbashian}.

\section{Other correlations}

Although we studied in detail only the two-point correlation function, the equilibration properties of any higher 
order ($n$-point) 
correlation function of the bosonic fields $\phi$ follow those of the two-point function, by virtue of Wick's theorem. 
In particular they are also given by their GGE expectation values. This result holds generally for quenches in free 
systems \cite{SC14}. 

Moreover, extending the results \cite{SC14} and anticipated in some form by \cite{CE-10}, we conclude that the 
same is true for any initial state that satisfies the cluster decomposition principle (i.e. any state in which all 
connected multi-point correlations of the fields decay at large distances), even if it does not satisfy the far more 
restricting property of Wick's theorem (i.e. that the connected correlations at finite distances are identically zero). 
For all such states, when they evolve under the long-range \emph{harmonic} Hamiltonian, any multi-point correlation 
function equilibrates at large times. In particular, this means that when the initial state is the ground state (but 
also an excited or a thermal state) of an \emph{anharmonic} long-range Hamiltonian, instead of a harmonic one, it also 
leads to equilibration, as long as it satisfies the cluster decomposition principle.

We can also consider the two-point correlation function of the canonical momentum field $\pi\equiv\dot\phi$. 
It can be readily seen that
\begin{align}
\frac{d^2}{dt^2} \left(\phi_{k} \phi_{-k}\right) &  = \ddot \phi_{k}  \phi_{-k} + \phi_{k} \ddot \phi_{-k} + 2 \pi_{k} \pi_{-k}  \nonumber \\
& = - 2 \omega^2(k) \phi_{k} \phi_{-k} + 2 \pi_{k} \pi_{-k} 
\end{align}
where in the last line we used the equations of motion $\ddot\phi_{k} = -\omega^2(k) \phi_{k}$. Therefore
\be
\pi_{k} \pi_{-k}  = \frac12 \frac{d^2}{dt^2} \left(\phi_{k} \phi_{-k}\right) + \omega^2(k) \phi_{k} \phi_{-k}
\ee
From the above relation, analogously to the expression for the $\phi$ correlation function 
Eq.~(\ref{Two point function after quench}), 
we can find that the two-point correlation function of $\pi$ is given by
\begin{align}
& F(r,t) \equiv \langle\pi(r,t)\pi(0,t)\rangle - \langle\pi(r,0)\pi(0,0)\rangle = \nonumber \\
& \int \frac{d^d k}{(2\pi)^d} e^{i\mathbf{k}.\mathbf{r}}
\frac{(\omega_0^2(k)-\omega^2(k))(\cos(2\omega(k)t)-1)}{\omega_0(k)}.
\end{align}
The large time behavior of the last expression can be studied in the same way as that of
(\ref{Two point function after quench}): 
it always becomes stationary, since for $m_0\neq 0$ its time independent part never exhibits 
IR divergences (only physically irrelevant 
UV divergences). Therefore we conclude that the $\pi$ correlations equilibrate for any values of 
the parameter $\alpha$ and of the masses.

\section{The effect of interactions}\label{sec:VI}

An interesting question is how our conclusions would be modified by the presence of interactions, i.e. 
if the Hamiltonian that describes the system after the quench was not quadratic but included anharmonic terms 
as
\begin{align} \label{anharmonic Hamiltonian real space omega1}
H_{\alpha}=&\frac{1}{2}\sum_r\pi^2(r)+\frac{1}{2}\sum_{r,r'}\phi(r)\Big{(}\frac{1}{|r-r'|^{d+\alpha}}+m^{\alpha}\delta_{rr'}\Big{)}\phi(r')\nonumber\\&+
\frac{\lambda}{4!}\sum_r\phi^4(r).
\end{align}

As far as equilibration is concerned, an answer to the  question posed at the beginning of the section can be induced 
by earlier studies based on 
self-consistent perturbation theory. In \cite{CS2010} it was shown that in the short-range case, the leading 
order effect of the interactions as given by a self-consistent approximation (Hartree-Fock or `loop' correction), 
is to shift the value of the mass to a higher effective value. Due to this shifting effect, equilibration is expected to 
remain unaffected by the inclusion of interactions, but also to extend to the massless case, since the zero mass 
would now be shifted to a positive effective value, allowing the system to equilibrate. This conclusion is expected 
to be correct in short-range systems at least in $2d$ and $3d$ where perturbation theory works fine also at equilibrium. 
Following this reasoning we could argue that the same is true also in the long-range case. Based on the
self-consistent approach of \cite{CS2010} one
finds that the effective mass at large times satisfies the equation
\begin{align} \label{effective mass 1}
&m^{*\alpha}=m^{\alpha}+\nonumber\\
&\frac{\lambda}{2}\int \frac{d^dk}{(2\pi)^d}\Big{(}\frac{(\omega_0(k)-\omega^*(k))^2}{4\omega_0(k)\omega^{*2}(k)}+
\frac{\omega(k)-\omega^*(k)}{2\omega(k)\omega^*(k)}\Big{)},
\end{align}
where $\omega^*(k)=\sqrt{|k|^{\alpha}+m^{*\alpha}}$. In principle, since in $d=2$ and $3$ the system equilibrates even when we do not
have interactions, based on the above argument one can simply conclude that this will be the case also 
when we do have interactions. The interesting case is the massless case in $d=1$ with $1\leq\alpha<2$ where, as we saw, in the absence of 
interactions there is no equilibration. 

After making  the integrals in the equation (\ref{effective mass 1}) dimensionless we have
\begin{align} \label{effective mass 2}
m^{*\alpha}=m^{\alpha}+\frac{\lambda}{2\pi}
\left[m_0^{1-\alpha/2}f_{\alpha}\left(\frac{m^*}{m_0}\right)+ {m^*}^{1-\alpha/2} h_{\alpha}\left(\frac{m}{m^*}\right)\right],
\end{align}
where
\begin{align} \label{f and g}
f_{\alpha}(s)\equiv&\int_0^{\infty}dk
\frac{(\sqrt{k^{\alpha}+1}-\sqrt{k^{\alpha}+s^{\alpha}})^2}{4\sqrt{k^{\alpha}+1}({k^{\alpha}+s^{\alpha}})} \\
h_{\alpha}(s) \equiv &\int_0^{\infty}dk
\frac{\sqrt{k^{\alpha}+s^{\alpha}}-\sqrt{k^{\alpha}+1}}{2 \sqrt{k^{\alpha}+s^{\alpha}} \sqrt{k^{\alpha}+1}} \nonumber\\
=& (1-s^{1-\alpha/2}) {\Gamma(1/\alpha) \Gamma(1/2-1/\alpha)} / ({2\sqrt{\pi} \alpha}).
\end{align}

\begin{figure}[h]
\begin{center}
\includegraphics[clip,width=\linewidth]{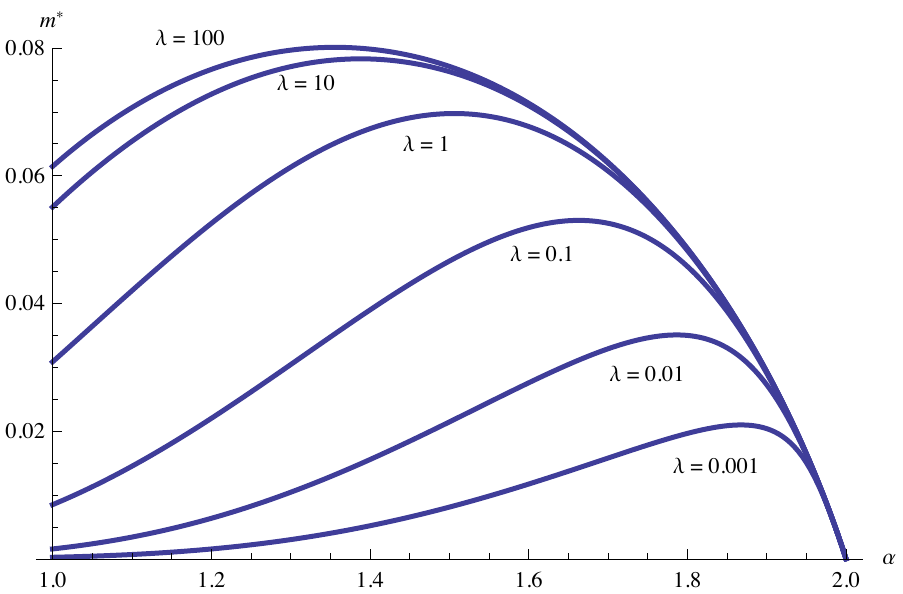}
\caption{\label{fig:interactions} Plots of the effective
mass $m^*$ for $m=0$ in one dimension as a function of $\alpha$ in the range from 1 to 2, for various values of $\lambda$ (0.001 to 100) 
in units of $m_0=1$. The results are obtained by numerical solution of the self-consistency equation (\ref{effective mass 2}). The solutions are real positive. Note also that the curves tend to zero in the short-range limit $\alpha=2$ with the same slope for all $\lambda$.}
\end{center}
\end{figure}

Solving numerically the equation (\ref{effective mass 2}) for $m=0$ and $1\leq\alpha<2$ we find that the solution $m^*$ is real and positive (non-zero). The results are shown in Fig.~\ref{fig:interactions} where the solutions $m^*$ are plotted as functions of $\alpha$ for various values of $\lambda$. This is a remarkable result because it actually tells us that in 1$d$, if we quench the system to zero mass and $1\leq\alpha<2$, according to the above approximation the interactions will produce a nonzero effective mass $m^*$, which will lead the system to equilibration. This is in contrast to what happens in the absence of interactions, where the system does not equilibrate for these parameter values.

In the above calculation, we did not include explicitly in the Hamiltonian (\ref{anharmonic Hamiltonian real space omega1}) a short-range kinetic term, i.e. $g(\nabla\phi(r))^2$, which appears anyway when one studies long-range spin systems \cite{Fisher}. 
In the last forty years different versions of renormalization group have been applied to the field theory $H_{g}=H_{\alpha}+g(\nabla\phi(r))^2$
with different contradicting conclusions \cite{Fisher,Sak,Picco-Rajabpour}. The main difference between these papers is that the short-range regime of the Ising model starts from $\alpha=2$ or $\alpha=2-\eta$, where $\eta$ is the anomalous dimension of 
spin operator (in $d=1$ it is equal to $\frac{1}{4}$). What all the studies agree on is that for $\alpha<2-\eta$ one can ignore the short-range 
kinetic part i.e. just set $g=0$ and do the renormalization group analysis \cite{eta}. In principle it is not obvious that this line
of reasoning can be applied also in the case of the quench problem. In non-equilibrium systems one would have to use an argument
from boundary renormalization group in a slab geometry \cite{CS2010} which in the case of long-range systems is completely unexplored in the literature. This is mainly because in the presence of long-range systems it is very difficult to control the boundary conditions. For this reason
we are unable to make any concrete prediction regarding the equilibration properties of the Hamiltonian $H_{g}$.

It is also important to mention implications of the above calculations on an experimentally relevant system of quantum long-range 
Ising model with the Hamiltonian:
\begin{align} \label{Hamiltonian LR Ising}
H=-\sum_{i<j} J_{ij}\sigma_i^z\sigma_j^z-h\sum_i \sigma_i^x,
\end{align}
where $J_{ij}=\frac{1}{|i-j|^{d+\alpha}}$. It is widely believed that the above system can be described 
with the Hamiltonian (\ref{anharmonic Hamiltonian real space omega1}) and different
phases can appear depending on the value of $\alpha$. For example in \cite{DB01} it is argued that for $\alpha<\frac{2d}{3}$ the system follows the mean-field 
behavior and so can be described with the Hamiltonian (\ref{Quadratic Hamiltonian}) with the dispersion relation Type-1. Based on our calculations
one can predict that the $\sigma^z$ operator is going to equilibrate after quantum quench independent of the value of $\alpha$. Another interesting
observation is that in the mean-field regime one might be able to construct some sort of GGE for the non-integrable Hamiltonian (\ref{Hamiltonian LR Ising}).
Finally it is also interesting to comment about the case $\alpha<0$. In principle the Hamiltonians (\ref{Quadratic Hamiltonian real space omega1}) and 
(\ref{Hamiltonian LR Ising}) can be also defined
for all the values of $\alpha$ with $-d<\alpha<0$. However, in this regime the Hamiltonian (\ref{Quadratic Hamiltonian}) has unusual behavior,  the modes with smaller $k$  have higher energy than the modes with bigger $k$. 
Ignoring the complications regarding the thermodynamic limit of these systems
see \cite{CDR2009} and references therein, one can simply follow the
same line of thinking that we followed for positive $\alpha$'s and conclude that the $\phi$ and $\pi$ operators are going to equilibrate also in this case.
At the moment it is not obvious for us that one can extend this conclusion also to the operators in the Hamiltonian (\ref{Hamiltonian LR Ising}) with $\alpha<0$.

To summarize this section, we found that equilibration happens also in the case of interacting long-range evolution in all 
dimensions and for all the values of $m$ and $\alpha$. In fact we saw that it happens even in the cases where it does not happen 
in the non-interacting case. Our approximation on which we based these results, is to map the interacting non-integrable model 
to a free and therefore integrable system, with an interaction dependent effective mass that is determined self-consistently. 
In this approximation the equilibration is described by the GGE with the mass replaced by its effective value. However, 
since the original interacting field theory that we considered is non-integrable, one would expect that after the emergence 
of this quasi-stationary state at intermediate times (\emph{prethermalization}), the interactions will slowly transfer energy
from one momentum mode to the other, so that the different effective temperatures of each mode will eventually converge to
the same value and true thermalization will take place. The validity of this scenario remains under investigation \cite{r-09,r-09a,prethermalization1,
prethermalization2,prethermalization3,prethermalization4,prethermalization5,prethermalization6,prethermalization7} with a 
few recent studies having observed prethermalization also in interacting long-range systems \cite{prethermalization8,prethermalization9}.

\section{Conclusions}

In this paper we studied the time evolution of correlations after a quantum quench of the mass parameter in long-range 
free field theory for different values of the power of the couplings $\alpha$ and for two different types of dispersion 
relations, in one, two and three spatial dimensions. We classified the parameter regimes in which the system equilibrates 
and those in which it exhibits a horizon effect. We found that in contrast to what happens in short-range systems, the 
horizon may be fake, in the sense that correlations outside of it do not decay exponentially, as is the case of the 
Lieb-Robinson bound.

We also studied the effect of the initial state in the dynamics of the correlations, more specifically of the type 
of the dispersion 
relation which affects the decay of the initial correlations with the distance. We showed that both the presence of 
equilibration and 
horizon effect remain unaffected by this change, however when the initial correlations decay exponentially instead 
of like a power-law, 
the horizon turns from fake to true. These results can be successfully explained by the semiclassical approach, 
which in general describes many aspects of quantum quench problems. Furthermore, we considered the effect 
of the interactions and, following earlier work, we concluded that the presence of interactions in either the
pre-quench or the post-quench Hamiltonian does not spoil but rather enhances the equilibration properties of the system.

Finally we showed that despite the fact that the integrals of motions in our system are non-local, in the regimes where 
equilibration occurs, one can still describe the long-time stationary values by a GGE.

\textit{Acknowledgments}

MAR's work was partially supported by FAPESP. SS's work was supported by the ERC under Starting Grant 279391 EDEQS. 
This work was initiated during MAR's visit to the Physics Department of the University of Pisa using the 
European fellowship IRSES Grant QICFT. We thank also Pasquale Calabrese, Mario Collura and 
Thiago Rodrigues de Oliveira  for fruitful comments.


\begin{thebibliography}{99}
\bibitem{eta} The basic idea of \cite{Sak} is that in the presence of interaction the $\phi$ operator has an anomalous dimension $\eta$ at $\alpha=2$.
If we now turn on the long-range kinetic term this term will be irrelevant up to $\alpha=2-\eta$. This reasoning is
in contradiction with the numerical results in \cite{Picco} and analytical results in \cite{Picco-Rajabpour}.
\end{thebibliography}
\end{document}